# Modeling Terahertz Propagation via Frequency-Domain Physics-Informed Neural Networks

Pengfei Zhu, *Graduate Student Member*, *IEEE*, Hai Zhang, *Member*, *IEEE*, Stefano Sfarra, Elena Pivarčiová, Cunlin Zhang, Xavier Maldague, *Life Senior Member*, IEEE

*Abstract*—Terahertz time-domain spectroscopy (THz-TDS) provides a non-invasive and label-free method for probing the internal structure and electromagnetic response of materials. Numerical simulation of THz-TDS can help understanding wave-matter interactions, guiding experimental design, and interpreting complex measurement data. However, existing simulation techniques face challenges in accurately modeling THz wave propagation with low computational cost. Additionally, conventional simulation solvers often require dense spatial-temporal discretization, which limits their applicability to large-scale and real-time scenarios. Simplified analytical models may neglect dispersion, multiple scattering, and boundary effects. To address these limitations, we establish a novel computational framework that integrates frequency-domain physics-informed neural networks (FD-PINNs) with less data-driven. To validate our proposed FD-PINNs, simulation results from finite-difference time-domain (FDTD) and time-domain (TD)-PINNs were used to compare with FD-PINNs. Finally, experimental results from THz-TDS systems were employed to further exhibit accurate reconstruction ability of FD-PINNs.

This work was supported in part by the Natural Sciences and Engineering Research Council of Canada (NSERC) through the CREATE-oN DuTy! Program (496439-2017) and the Discovery Grants Program (RGPIN-2020-04595), and in part by the Canada Research Chair in Multi-polar Infrared Vision (MIVIM). Corresponding authors: Hai Zhang.

Pengfei Zhu, Hai Zhang and Xavier Maldague are with the Department of Electrical and Computer Engineering, Computer Vision and Systems Laboratory (CVSL), Laval University, Québec G1V 0A6, Québec city, Canada (e-mail: pengfei.zhu.1@ulaval.ca; hai.zhang.1@ulaval.ca; xavier.maldague@gel.ulaval.ca). Hai Zhang is also with the Beijing Advanced Innovation Center for Imaging Technology and Key Laboratory of Terahertz Optoelectronics (MoE) and the Centre for Composite Materials and Structures (CCMS), Harbin Institute of Technology, Harbin 150001, China (hai.zhang@hit.edu.cn).

Stefano Sfarra is with the Department of Industrial and Information Engineering and Economics (DIIIE), University of L'Aquila, I-67100 L'Aquila, Italy (stefano.sfarra@univaq.it)

Elena Pivarčiová is with the Faculty of Environmental and Technology Manufacturing, Technical University in Zvolen, 960 53, Zvolen, Slovakia (pivarciova@tuzvo.sk)

Cunlin Zhang is with the Beijing Advanced Innovation Center for Imaging Technology and Key Laboratory of Terahertz Optoelectronics (MoE), Department of Physics, Capital Normal University, Beijing 100048, China (cunlin_zhang@mail.cnu.edu.cn)

## I. INTRODUCTION

THE interest in terahertz (THz) technology, sparked by the many possible applications such as communications [1], [2], spectroscopy [3], [4], and imaging [5], [6], has given rise to the emergence of novel experimental and simulation methods for terahertz time-domain spectroscopy (THz-TDS) [7] and near-field scanning techniques [8]. However, due to strong dispersion and interface effect in THz band and limited computational method, the simulation of THz propagation faces challenge.

There are two mainstream numerical methods for simulating the electromagnetic (EM) wave propagation: the finite-difference time-domain (FDTD) method and the finite element method (FEM). FDTD is widely used for modeling EM pulse propagation in complex media [9], [10], and has been extensively applied to study nonlinear magnetic and electric response in various materials [11], [12]. FEM, on the other hand, provides a versatile framework for solving EM problems in both the time and frequency domains [13], [14]. Owing to its flexibility, FEM has been successfully employed to investigate EM propagation across a wide spectral range, from microwave to optical frequencies [15], [16], making it a natural candidate for terahertz studies as well. Despite their advantages, both FDTD and FEM face a stringent spatial resolution requirement: the maximum element size should typically be smaller than one-third of the wavelength, and in many high-fidelity simulations, reduced further to one-tenth, to ensure numerical stability and accuracy [17].

In wireless communications fields, channel modelling in THz band is envisioned as a key technology to support ultra-broadband wireless systems for beyond 5G. To analyze very large structures with reasonable computational complexity [18], ray-tracing (RT) has emerged as a popular technique for the analysis of site-specific scenarios [19]. However, it has less accuracy for fine structures and complex materials and is difficult to capture wave phenomena such as interference and fine diffraction effects. Some researcher proposed the statistical methods, which use statistical models, such as the Saleh-Valenzuela (SV) model [20], geometry-based stochastic channel model (GSCM) [21], and Gaussian mixture model (GMM) [22], to approximate complex multipath propagation, avoiding the need for precise modeling of each scatterer.

Although the statistical methods have high computational efficiency without full geometric details, they still lack sufficient accuracy. Hybrid methods, such as stochastic scatter placement and RT hybrid approach (SSRTH) [23] and RT-FDTD hybrid modeling [24], combine deterministic and statistical methods for balancing accuracy and computational efficiency. Nevertheless, complex implementation, parameter design, and problem of boundary transitions significantly limit the further development of hybrid methods.

Physics-informed neural networks (PINNs) can seamlessly integrate experimental data with the various partial differential equations (PDEs) for solid/fluid mechanics [25], [26], quantum mechanics [27], heat transfer [28], etc. In the context of EM wave propagation at THz frequencies, PINNs represent a promising alternative to conventional numerical solvers such as FDTD and FEM. Unlike mesh-based approaches, PINNs incorporate the underlying physical laws directly into the neural network's loss function, allowing the PDEs to be solved without explicit meshing. This mesh-free formulation not only reduces the complexity associated with discretizing intricate geometries but also enables efficient handling of heterogeneous media and high-dimensional problems. Additionally, the integration of experimental or boundary data into the training process enhances the model's predictive accuracy and generalization capabilities, making PINNs particularly suitable for challenging THz wave propagation scenarios.

The first attempt of applying PINNs to THz fields is performed by Zhu et al. [29]. The Maxwell equations of describing THz propagation are simplified as wave equations. According to the previous cases, PINNs are difficult to fit wave problem, even if a simple one-dimensional wave equation [30]. Additionally, ultrafast time scale and low photonic energy also hinder us to simulate THz propagation.

In this work, we proposed the terahertz-frequency-domain physics-informed neural networks (THz-FD-PINNs) for modeling THz wave propagation, for the first time. THz-FD-PINNs can directly simulate the frequency-domain spectrum with low computational cost. Furthermore, the time-domain (TD)-PINNs is also employed to validate the accuracy of proposed THz-FD-PINNs. To accelerate the training of THz-FD-PINNs, a conjugate symmetry-based truncated strategy is used for FD-PINNs. Both simulations and experiments are conducted to exhibit the feasibility of proposed THz-FD-PINNs, where simulation data comes from FDTD while experimental data comes from THz-TDS systems.

II. THEORY OF TERAHERTZ PHYSICS-INFORMED NETWORKS

The Maxwell equations can be used to describe the THz wave propagation. In this work, the frequency-domain Maxwell equations can be given as:

$$\nabla \cdot \mathbf{D}(\mathbf{r}, \omega) = \rho(\mathbf{r}, \omega) \quad (1)$$
$$\nabla \cdot \mathbf{B}(\mathbf{r}, \omega) = 0 \quad (2)$$
$$\nabla \times \mathbf{E}(\mathbf{r}, \omega) = -i\omega \mathbf{B}(\mathbf{r}, \omega) \quad (3)$$
$$\nabla \times \mathbf{H}(\mathbf{r}, \omega) = \mathbf{J}(\mathbf{r}, \omega) + i\varepsilon \mathbf{D}(\mathbf{r}, \omega) \quad (4)$$

where $\mathbf{D}(\mathbf{r}, \omega) = \epsilon(\omega)\mathbf{E}$ denotes the electric flux density, $\mathbf{E}$ denotes electric field vector, $\mathbf{H}$ denotes magnetic field strength vector, $\mathbf{B}(\mathbf{r}, \omega) = \mu(\omega)\mathbf{H}$ denotes the magnetic flux density, $\rho$ denotes the free charge density, $\mathbf{J}(\mathbf{r}, \omega) = \sigma(\omega)\mathbf{E}$ denotes the electric current density vector, $\epsilon$ denotes the permittivity of the medium, $\mu$ denotes the permeability of the medium, $\sigma$ denotes the electrical conductivity of the medium, and $t$ denotes the time. The above equations can be arranged as:

$$\nabla^2 \mathbf{E} - \nabla(\nabla \cdot \mathbf{E}) + \omega^2 \mu \tilde{\epsilon} \mathbf{E} = 0 \quad (5)$$

where $\tilde{\epsilon} = \epsilon - i\frac{\sigma}{\omega}$ is the relative permittivity. Assuming there is no free charge and current, i.e., $\rho = 0$, the equation is degraded as:

$$\nabla^2 \mathbf{E} + \omega^2 \mu \tilde{\epsilon} \mathbf{E} = 0 \quad (6)$$

where $\mu = 4\pi \times 10^{-7}$ is the vacuum magnetic permeability.

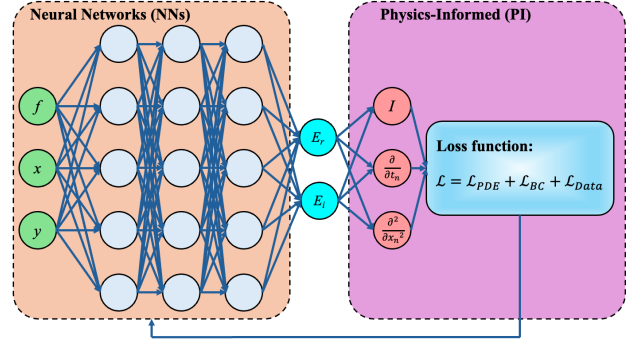

Fig. 1. Schematic of THz-FD-PINNs.

The structure of terahertz frequency-domain physics-informed neural networks (THz-FD-PINNs) is shown in Fig. 1. It has been validated that frequency domain PDE loss is better than time-domain PDE loss in heat transfer problem [28]. The complete loss function in THz-FD-PINNs can be given as:

$$\mathcal{L} = \mathcal{L}_{PDE} + \mathcal{L}_{BC} \quad (7)$$

where $\mathcal{L}_{PDE}$ is the loss function for PDE, and $\mathcal{L}_{BC}$ is the loss function for boundary condition. Of note, there is no initial condition in THz-FD-PINNs. This is because the time-domain initial conditions are encoded into the amplitude and phase of the frequency domain fields through the Fourier transform. The $\mathcal{L}_{PDE}$ is formulated as:

$$\mathcal{L}_{PDE} = \frac{1}{N_f}\sum_{j=1}^{N_f}(\nabla^2 \mathbf{E} - \nabla(\nabla \cdot \mathbf{E}) + \omega^2 \mu \tilde{\epsilon} \mathbf{E})^2 \quad (8)$$

where $n$ is the refractive index of the medium, and $N_f$ is the number of data points for PDE terms. Different from time-domain (TD)-PINNs, the PDE loss functions of THz-FD-PINNs include two parts, i.e., real part and imaginary part:

$$\mathcal{L}_{PDE} = \mathcal{L}_{real} + \mathcal{L}_{imag} \quad (9)$$

$$\mathcal{L}_{real} = \frac{1}{N_f}\sum_{j=1}^{N_f}(\nabla^2 \mathbf{E}_r - \nabla(\nabla \cdot \mathbf{E}_r) + \omega^2\mu(\epsilon \mathbf{E}_r + \frac{\sigma}{\omega}\mathbf{E}_i))^2 \quad (10)$$

$$\mathcal{L}_{imag} = \frac{1}{N_f}\sum_{j=1}^{N_f}(\nabla^2 \mathbf{E}_i - \nabla(\nabla \cdot \mathbf{E}_i) + \omega^2\mu(\epsilon \mathbf{E}_i - \frac{\sigma}{\omega}\mathbf{E}_r))^2 \quad (11)$$

where $\mathbf{E}_r$ and $\mathbf{E}_i$ are the real part and imaginary part of $\mathbf{E}$. For THz-TDS, there is an incident light source $\mathbf{f}$. Therefore, the boundary conditions can be given as:

$$\mathcal{L}_{BC} = \frac{1}{N_{BC}}\sum_{j=1}^{N_{BC}}[(\mathbf{E}_r - \mathbf{f}_r)^2 + (\mathbf{E}_i - \mathbf{f}_i)^2] \quad (12)$$

where $\mathbf{E}_r$ and $\mathbf{E}_i$ are the real part and imaginary part of $\mathbf{f}$, and $N_{BC}$ is the number of data points for boundary condition terms One of the PINNs' advantages is that the incident source can be random function or user definition.

Additionally, PINNs can incorporate pre-known data into the network, which accelerates convergence and improves training efficiency. Accordingly, an additional data-driven loss term is introduced, constraining the network output to match the response governed by the Maxwell equations under investigation:

$$\mathcal{L}_{Data} = \frac{1}{N_{Data}}\sum_{j=1}^{N_{Data}}[(\mathbf{E}_r - \mathbf{E}_{d\_r})^2 + (\mathbf{E}_i - \mathbf{E}_{d\_i})^2] \quad (13)$$

where $N_{Data}$ is the number of data points for data loss $\mathcal{L}_{Data}$, $\mathbf{E}_{d\_r}$ and $\mathbf{E}_{d\_i}$ are the real part and imaginary part of data points $\mathbf{E}_d$. All models were implemented with the PyTorch framework, a code database in Python software, specifically developed for achieving deep learning, and then trained using NVIDIA 4060 Titan GPUs. The hyperbolic tangent (Tanh) is selected as the activation function. The network consists of eight layers: one input layer, six hidden layers, and one output layer, with each hidden layer containing 128 neurons. To balance the contribution of different loss terms in training the PINNs, we employ a gradient-based adaptive weighting scheme. Specifically, for each loss term $\mathcal{L}_k$, we compute its gradient norm $g_k = ||\nabla_\theta \mathcal{L}_k||$ and assign the weight $\lambda_k = \frac{1/g_k}{\sum_j 1/g_j}$. This approach amplifies loss terms with small gradients, ensuring that they are not overlooked during training. The total epoch is set to 50000. The Adam optimizer is used with $1 \times 10^{-3}$ learning rate.

## III. EXPERIMENTS AND SIMULATION

### A. Experimental Setup

The configuration of the THz imaging system is illustrated in Fig. 2(a). An ultrafast laser pulse is divided into a pump beam and a reference beam. The pump beam is delayed by an optical time-delay line and directed to a THz emitter to generate a THz pulse. The emitted THz wave propagates through the sample and is detected by a coupled detector, while the reference beam serves as the sampling signal. The sampled THz signal is subsequently processed by a lock-in amplifier to enhance the weak signal for data acquisition. The experimental system was provided by Menlo Systems GmbH (Munich, Germany) with a frequency resolution of 1.2 GHz and a repetition rate of 100 MHz. Measurements were conducted in both transmission and reflection modes, with a scanning step size of 0.5 mm. Of note, in THz-TDS results, we only obtained THz signals from the external surface, whether in transmission or reflection mode.

### B. Numerical Simulation

In a two-dimensional Cartesian coordinate system, the finite-difference time-domain (FDTD) formulation is obtained by substituting central-difference approximations for the spatial and temporal derivatives in Maxwell's curl equations [31], [32], [33], [34]. This procedure yields a set of coupled time-stepping relations for the electric- and magnetic-field components. For a 2D model, the update equations can be expressed as

$$\left(1 + \frac{\Delta t \cdot \sigma}{2\varepsilon}\right)E_x^{n+1}\left(i + \frac{1}{2}, j\right) = \left(1 - \frac{\Delta t \cdot \sigma}{2\varepsilon}\right)E_x^n\left(i + \frac{1}{2}, j\right) + \frac{\Delta t}{\delta\varepsilon}[H_z^{n+\frac{1}{2}}(i + \frac{1}{2}, j + \frac{1}{2}) - H_z^{n+\frac{1}{2}}(i + \frac{1}{2}, j - \frac{1}{2})] \quad (14)$$

$$\left(1 + \frac{\Delta t \cdot \sigma}{2\varepsilon}\right)E_y^{n+1}\left(i, j + \frac{1}{2}\right) = \left(1 - \frac{\Delta t \cdot \sigma}{2\varepsilon}\right)E_y^n\left(i, j + \frac{1}{2}\right) - \frac{\Delta t}{\delta\varepsilon}[H_z^{n+\frac{1}{2}}(i + \frac{1}{2}, j + \frac{1}{2}) - H_z^{n+\frac{1}{2}}(i - \frac{1}{2}, j + \frac{1}{2})] \quad (15)$$

$$\left(1 + \frac{\Delta t \cdot \sigma}{2\varepsilon}\right)H_x^{n+1/2}\left(i, j + \frac{1}{2}\right) = \left(1 - \frac{\Delta t \cdot \sigma}{2\varepsilon}\right)H_x^{n-1/2}\left(i, j + \frac{1}{2}\right) + \frac{\Delta t}{\delta\varepsilon}[E_y^n(i, j + \frac{1}{2}) - E_y^n(i, j)] \quad (16)$$

$$\left(1 + \frac{\Delta t \cdot \sigma}{2\varepsilon}\right)H_y^{n+1/2}\left(i + \frac{1}{2}, j\right) = \left(1 - \frac{\Delta t \cdot \sigma}{2\varepsilon}\right)H_y^{n-1/2}\left(i + \frac{1}{2}, j\right) - \frac{\Delta t}{\delta\varepsilon}[E_x^n(i + \frac{1}{2}, j) - E_x^n(i, j)] \quad (17)$$

where $i$ and $j$ denote the spatial indices of the Yee grid, and $n$ denotes the temporal index. Equations (14) and (15) describe the updates for the electric-field components, which are defined along the edges of each Yee cell, whereas (16) and (17) describe the magnetic-field components, which are defined at the cell faces. In addition to the half-cell spatial staggering between $E$ and $H$, the scheme incorporates a half-time-step staggering, which ensures second-order accuracy in both space and time. The simulation setup is shown in Fig. 2(b). The simulation duration was set to 30 ps. A plane-wave light source was injected into the material with a central frequency of 1 THz, a pulse length of 0.5 ps, a temporal offset of 2 ps, and a bandwidth of 1.5 THz.

### C. Materials

Generally, optical parameter measurements are performed in transmission to minimize the influence of alignment errors. The refractive index $n(\omega)$ and absorption coefficient $\alpha(\omega)$ can then be determined from the phase difference between the sample and reference signals [35], [36]:

$$n(\omega) = 1 + \frac{c}{2\pi\omega d}(\phi_s(\omega) - \phi_r(\omega)) \quad (18)$$

$$\alpha(\omega) = -\frac{2}{d}\ln[r(\omega)\frac{(n(\omega)+1)^2}{4n(\omega)}] \quad (19)$$

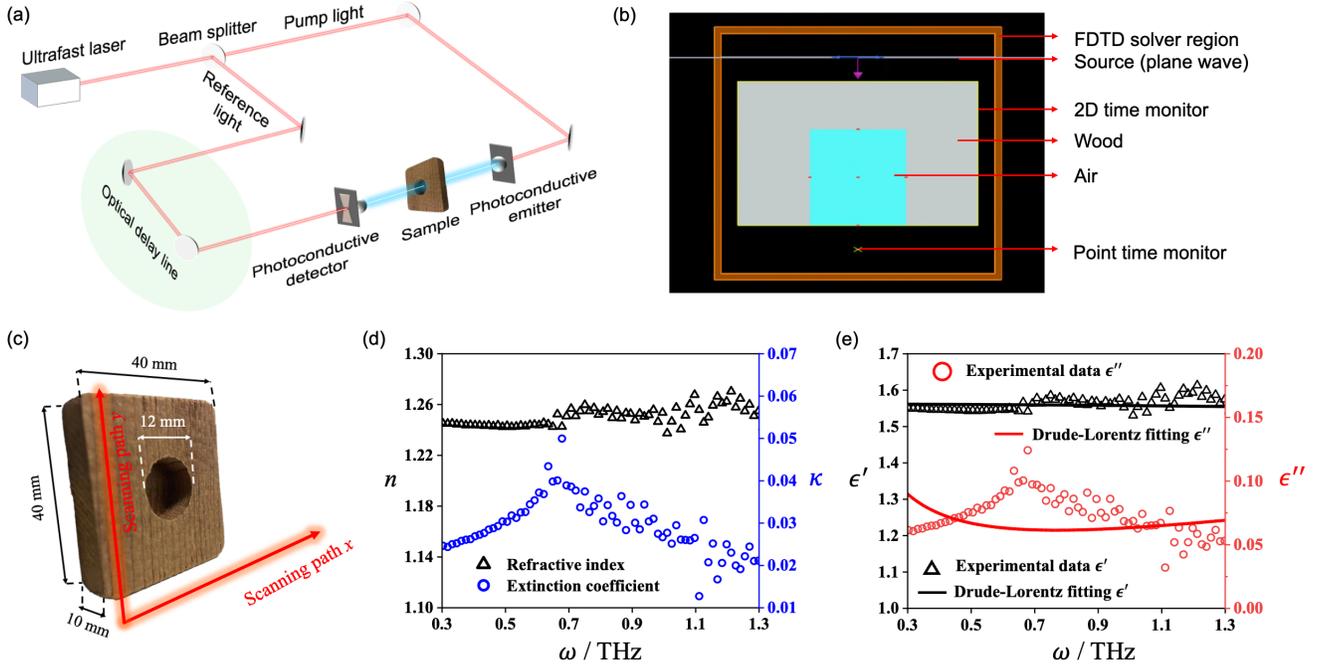

Fig. 2. Experiments and simulation for THz-TDS: (a) Schematical image of THz-TDS systems; (b) Schematical image of FDTD simulation; (c) Photograph of the sample; (d) Refractive index and extinction coefficient of the sample; (e) Real and imaginary parts of permittivity based on experimental data and Drude-Lorentz fitting.

where $\phi_s(\omega)$ and $\phi_r(\omega)$ denote the phase angles of the sample and reference signals, $c$ is the speed of light, $\omega$ is the frequency, and $d$ is the sample thickness.

The electromagnetic dispersion effect describes the frequency-dependent variation of a material's complex permittivity, $\tilde{\epsilon} = \epsilon'(\omega) - i\epsilon''(\omega)$, where $\epsilon'(\omega)$ and $\epsilon''(\omega)$ represent the real and imaginary parts, respectively. These quantities can be expressed as:

$$\tilde{\epsilon}(\omega) = [\tilde{n}(\omega)]^2 = [n(\omega) - i\kappa(\omega)]^2 \quad (20)$$
$$\epsilon'(\omega) = n^2(\omega) - \kappa^2(\omega) \quad (21)$$
$$\epsilon''(\omega) = 2n(\omega)\kappa(\omega) \quad (22)$$

where $\kappa(\omega) = \frac{\alpha(\omega)c}{4\pi\omega}$ is the extinction coefficient. In the THz spectral regime, the dispersive response of a material is primarily governed by both free-carrier contributions and bound-electron resonances. To accurately model this behavior, the complex permittivity can be described using a multi-oscillator Drude-Lorentz model:

$$\varepsilon(\omega) = \varepsilon_\infty - \frac{\omega_p^2}{\omega^2 + i\gamma\omega} + \sum_{j=1}^{N} \frac{\Delta\varepsilon_j \omega_{0,j}^2}{\omega_{0,j}^2 - \omega^2 - i\Gamma_j \omega} \quad (23)$$

where $\varepsilon_\infty$ is the high-frequency dielectric constant, $\omega_p$ is the plasma frequency, $\gamma$ is the damping factor, $\omega_{0,j}$ is the resonance frequency of the $j$-th oscillator, $\Delta\varepsilon_j$ is the strength of the $j$-th resonance, $\Gamma_j$ is the damping coefficient of the $j$-th resonance.

In this work, the experimental sample is a spruce wood with flat-bottom hole, as shown in Fig. 2(c). The size of the wood is 40 mm × 40 mm × 10 mm. The diameter and depth of flat-bottom hole are 12 mm and 7.5 mm. The material in FDTD includes lossless material and dispersive material. The refractive index of lossless material was set to 1.6. The material properties of dispersive material come from experimental results of the spruce wood. The refractive index and absorption coefficient were calculated based on equation (18) and (19), as shown in Fig. 2(d). The permittivity of experimental results was calculated based on equation (20)-(22), as shown in Fig. 2(e). Then, the Drude-Lorentz model was used to fit the experimental data based on equation (23) and fed into FDTD and THz-FD-PINNs, as shown in Fig. 2(d) and (e).

IV. RESULTS AND DISCUSSION

A. Non-Dispersive Medium

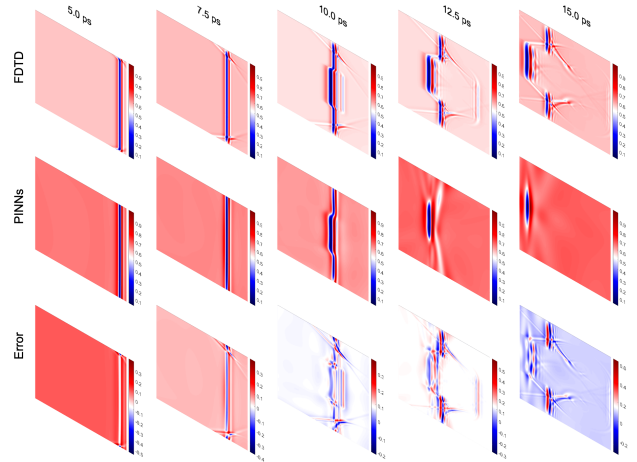

Fig. 3. Simulation results of FDTD and (time-domain) PINNs for non-dispersive medium.

The THz (time-domain) PINNs and THz-FD-PINNs were employed to simulate a simple example, i.e., non-dispersive medium. The principle of the time-domain PINNs is in the Appendix. For the original simulation setup in FDTD solvers, refractive index of materials was set to a constant value, 1.6. While the extinction value was set to zero. As mentioned before, few (20%) simulation data was used to accelerate the training process of PINNs and THz-FD-PINNs.

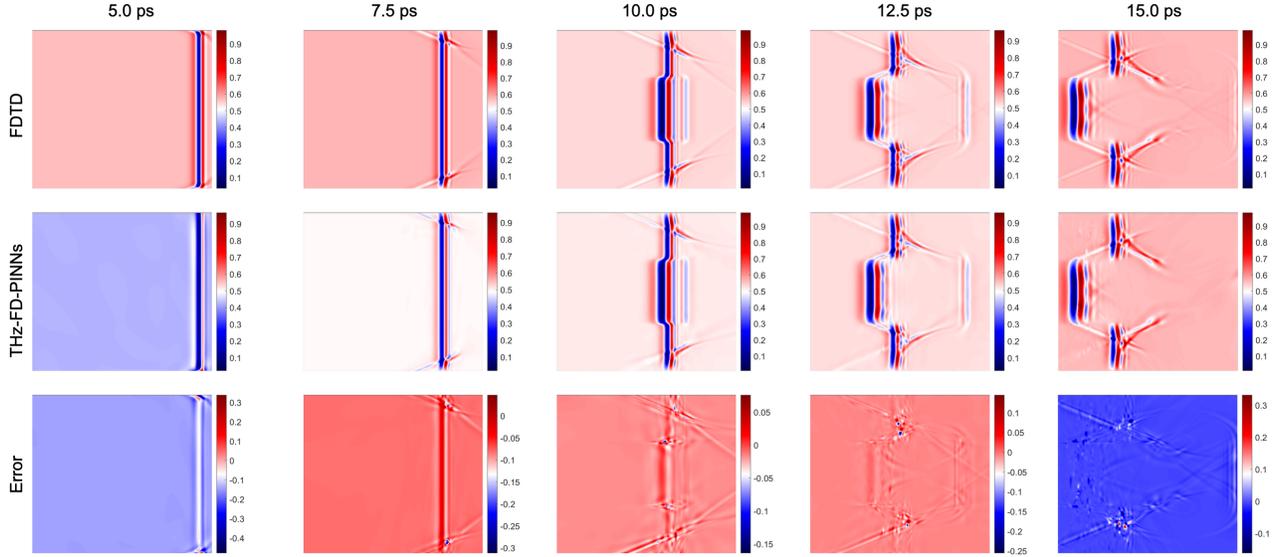

Fig. 4. Simulation results of FDTD and THz-FD-PINNs for non-dispersive medium.

The simulation results of (time-domain) PINNs and FDTD are shown in Fig. 3. At 5.0 ps and 7.5 ps, PINNs can effectively simulate the propagation of THz waves, although it cannot identify the scattering from sample's boundary. At 10.0 ps, THz wave meets the interface between the sample and the air. It appears obvious reflective and transmissive waves in FDTD results. However, the reflective wave in PINNs is not clear. At 12.5 ps, there is large difference between PINNs and FDTD results. We can only observe the transmissive wave in the air. Similar phenomenon exists at 15.0 ps. The error images denote the absolute error between FDTD and PINNs results. They are employed to demonstrate the prediction performance of PINNs.

The simulation results between THz-FD-PINNs and FDTD are shown in Fig. 4. At the initial stage (5.0 ps and 7.5 ps), the background magnitude of electromagnetics (EM) wave has relatively large difference with FDTD results. Nevertheless, it is clear to find the scattering phenomenon from sample's boundary at 7.5 ps. At 10.0 ps, THz-FD-PINNs can accurately simulate the reflective and transmissive waves caused by the interface. Furthermore, THz-FD-PINNs also predict the subsequent propagation of reflective and transmissive waves. It is obvious to find the delay effect of transmissive waves caused by different refractive indices between the sample and the air.

TABLE I
The Evaluation for PINNs and THz-FD-PINNs Using Root Mean Square Error.

| Time (ps) | PINNs | THz-FD-PINNs | Improvement |
|---|---|---|---|
| 5.0 | 7.78% | 14.64% | -6.86% |
| 7.5 | 5.33% | 8.21% | -2.88% |
| 10.0 | 7.18% | 1.81% | +5.37% |
| 12.5 | 12.64% | 1.11% | +11.53% |
| 15.0 | 10.52% | 1.10% | +9.42% |

To quantitatively compare the prediction accuracy between the time-domain PINNs and THz-FD-PINNs, we select the root mean square error as the evaluation index:

$$\text{RMSE} = \sqrt{\frac{1}{MN}\sum_{x=1}^{M}\sum_{y=1}^{N}(I(x,y)-\hat{I}(x,y))^2} \quad (24)$$

where $M$ and $N$ are the total pixel numbers along the length $x$ and width $y$ directions, respectively. $I$ and $\hat{I}$ are simulation results from FDTD and PINNs / THz-FD-PINNs. The RMSE values at different time are shown in Table 1. From the RMSE values at 5.0 and 7.5 ps, it is possible to find that PINNs are better than THz-FD-PINNs although PINNs cannot capture the boundary scattering effect. After 10.0 ps, THz-FD-PINNs are obviously better than PINNs. This result can be attributed by the influence from initial conditions. As we all know, in the time-domain PINNs, the initial conditions are strong constraints. However, in THz-FD-PINNs, there are no explicit initial conditions.

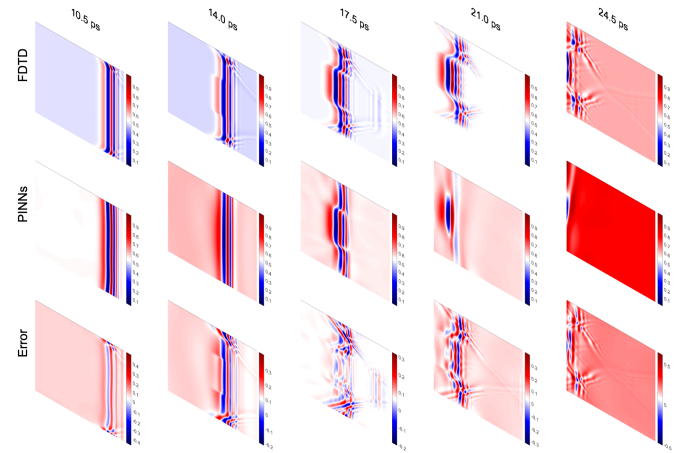

Fig. 5. Simulation results of FDTD and (time-domain) PINNs for dispersive medium.

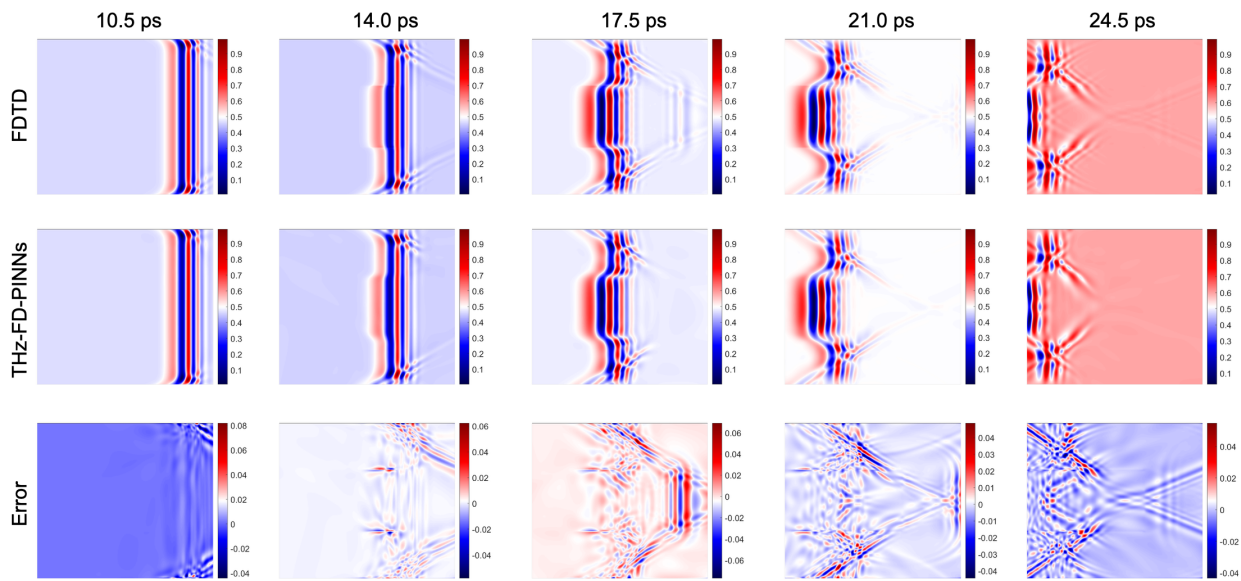

Fig. 6. Simulation results of FDTD and THz-FD-PINNs for dispersive medium.

## B. Dispersive Medium

After validating the excellent simulation performance of THz-FD-PINNs in non-dispersive medium, it is possible to explore the THz-FD-PINNs in dispersive medium. Actually, it is more meaningful to study the THz propagation in dispersive medium than that in non-dispersive medium, although most of existed literatures didn't consider the dispersive phenomenon in THz-TDS simulation. As we all know, THz wave's bandwidth locates at millimeter level, which can excite strong low-frequency vibration / rotation within molecules. This is also the reason why most of matter has specific fingerprint spectrum in THz waveband.

The simulation results of PINNs and FDTD are shown in Fig. 5. As mentioned before, the material parameters (permittivity) come from the Drude-Lorentz model (see Fig. 2(d) and (e)). Due to dispersive phenomenon in materials, PINNs cannot effectively simulate the background magnitude of EM field and interface effect. Especially for time at 21.0 ps and 24.5 ps, the prediction results of PINNs completely fails. Both reflective and transmissive waves have significant divergence with FDTD results.

The simulation results of THz-FD-PINNs and FDTD are shown in Fig. 6. It is possible to find that THz-FD-PINNs accurately simulate the propagation of THz waves whether it be transmission, reflection, or scattering phenomenon. According to error images, the maximum absolute error is no more than 0.08.

TABLE II
The Evaluation for PINNs and THz-FD-PINNs Using Root Mean Square Error.

| Time (ps) | PINNs | THz-FD-PINNs | Improvement |
| --- | --- | --- | --- |
| 10.5 | 6.08% | 3.10% | +2.98% |
| 14.0 | 11.69% | 2.90% | +8.79% |
| 17.5 | 5.76% | 0.71% | +5.05% |
| 21.0 | 7.03% | 0.46% | +6.57% |
| 24.5 | 18.07% | 0.53% | +17.54% |

The RMSE is employed to evaluate the performance between PINNs and THz-FD-PINNs, as shown in Table 2. It is possible to find that the THz-FD-PINNs significantly excels PINNs from the initial stage to the end stage. Especially for the later stage, PINNs are completely failed while THz-FD-PINNs remain high performance. Therefore, we can conclude that THz-FD-PINNs can effectively capture the spatial texture during THz propagation including slight scattering, transmission, reflection, and interface effect. However, due to the constrains from initial conditions, PINNs often exhibit higher accuracy at the beginning of THz propagation in the simple situation such as non-dispersive medium. Of note, due to introducing additional loss functions (real / imaginary loss), inevitably, THz-FD-PINNs often cost higher computational costs. In this work, for the same GPUs, THz-FD-PINNs have approximately twice the computation time than PINNs.

## C. Porous and Dispersive Medium

In the previous sections, we validated the excellent performance of THz-FD-PINNs for simulation. This type of sample often occurs in non-destructive testing fields. It is necessary to test the simulation performance of THz-FD-PINNs in a more complex sample, such as the porous and dispersive medium. Woods are the typical porous materials. In this work, the spruce wood with different size holes was studied, as shown in Fig. 7. There are three sizes of holes, including 0.3 mm, 0.2 mm, and 0.1 mm. The simulation setting and material parameters are the same to the previous section. The simulation results in time domain and frequency domain are shown in Fig. 7(b).

THz-FD-PINNs were employed to simulate the THz wave propagation in the porous and dispersive medium. The total epoch, hyperparameters, gradient-based adaptive weighting scheme, optimizer are the same to the previous training. The simulation results of THz-FD-PINNs and FDTD are shown in Fig. 7(c). Comparing with the results in Fig. 3 to Fig. 6, the scattering and diffraction effects become extremely complex in the porous and dispersive medium, especially in the time interval after 14.0 ps shown in the Fig. 7(c). However, THz-FD-PINNs still exhibit strong robustness and accuracy. The maximum error in all prediction results is no more than 0.35. Even though the intensity of electromagnetic wave becomes significantly weak caused by multiple refractions and scattering at 24.5 ps, the maximum error is no more than 0.15. These results demonstrate strong simulation capability and robustness of THz-FD-PINNs.

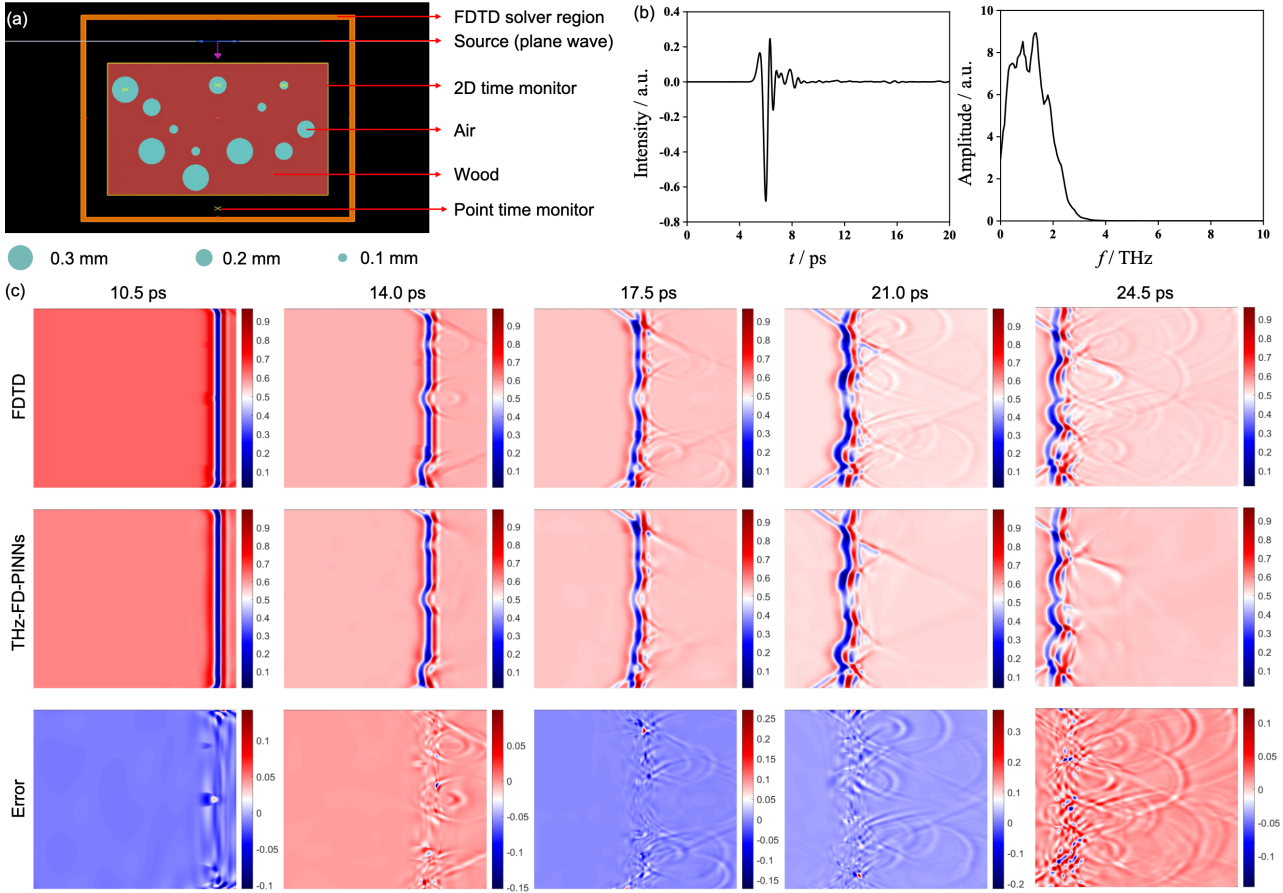

Fig. 7. Simulation results of FDTD and THz-FD-PINNs for porous and dispersive medium.

*D. Experimental Results*

After validating the excellent simulation performance of THz-FD-PINNs in both non-dispersive and dispersive medium, it is possible to test the prediction accuracy in real cases. As mentioned in Fig. 2, we scanned the spruce wood using THz-TDS systems. Besides the obtained optical parameters, we also have the experimental results of the 3D THz spectrum (*x*-*y*-*t*). Different from the previous simulation, the experimental results are captured from the back side of the sample in transmission mode. In other words, we only know results from the left side from Fig. 3 to Fig. 7 varying with time.

The experimental results of THz-TDS, time-domain PINNs, and THz-FD-PINNs are shown in Fig. 8. We select different profile lines around the center of woods, including the central position (0), shift right of the center (3.5 mm and 7 mm), shift left of the center (-3.5 mm and -7 mm). It is clear that the hole's sizes decrease with the deviation of the center. In addition, we compared the results from PINNs and THz-FD-PINNs. The simulation setting is similar to the previous sections. The only difference is that we input the 30% left boundary data for training. In contrast, we input the 30% all (internal and boundary) data for training in the previous section.

According to results from PINNs and THz-TDS, we can find that PINNs are not stable. There are significant deviations in 3.5 mm and -3.5 mm situations. Additionally, PINNs are difficult to capture the details of THz wave propagation, such as the secondary echoes. In contrary, THz-FD-PINNs exhibit excellent prediction capability in all experimental results. The main deviations come from the boundary of holes. This is due to the fact that we pre-set a constant value for each pixel's depth, and the light source is assumed as the plane source. In real THz-TDS experiments, the light source is a point light source, and the scattering phenomenon around the hole's boundary always exists.

## V. CONCLUSION

In this work, a novel simulation modality, physics-informed neural networks (PINNs) with less data-driven, was introduced for simulating the THz wave propagation, for the first time. Conventional simulation solvers such as FDTD and FEM rely on fine meshes and discretization, which lead to rapidly increasing computational cost with spatial and temporal dimensions. In contrast, PINNs embed Maxwell's equations directly into the loss function of a neural network, avoiding rigid mesh partitioning and enabling continuous-space modeling. However, the time-domain PINNs often face challenge to the problem of boundary scattering and interface effect, which significantly affect the simulation accuracy. To solve this problem, the terahertz frequency-domain physics-informed neural networks (THz-FD-PINNs) were proposed, for the first time. By multiple testing in non-dispersive, dispersive, and even porous-dispersive medium, THz-FD-PINNs exhibit high prediction accuracy and strong robustness. Finally, we employed the THz-FD-PINNs to the experimental data from THz-TDS systems. The results also demonstrate excellent simulation and prediction capabilities of THz-FD-PINNs.

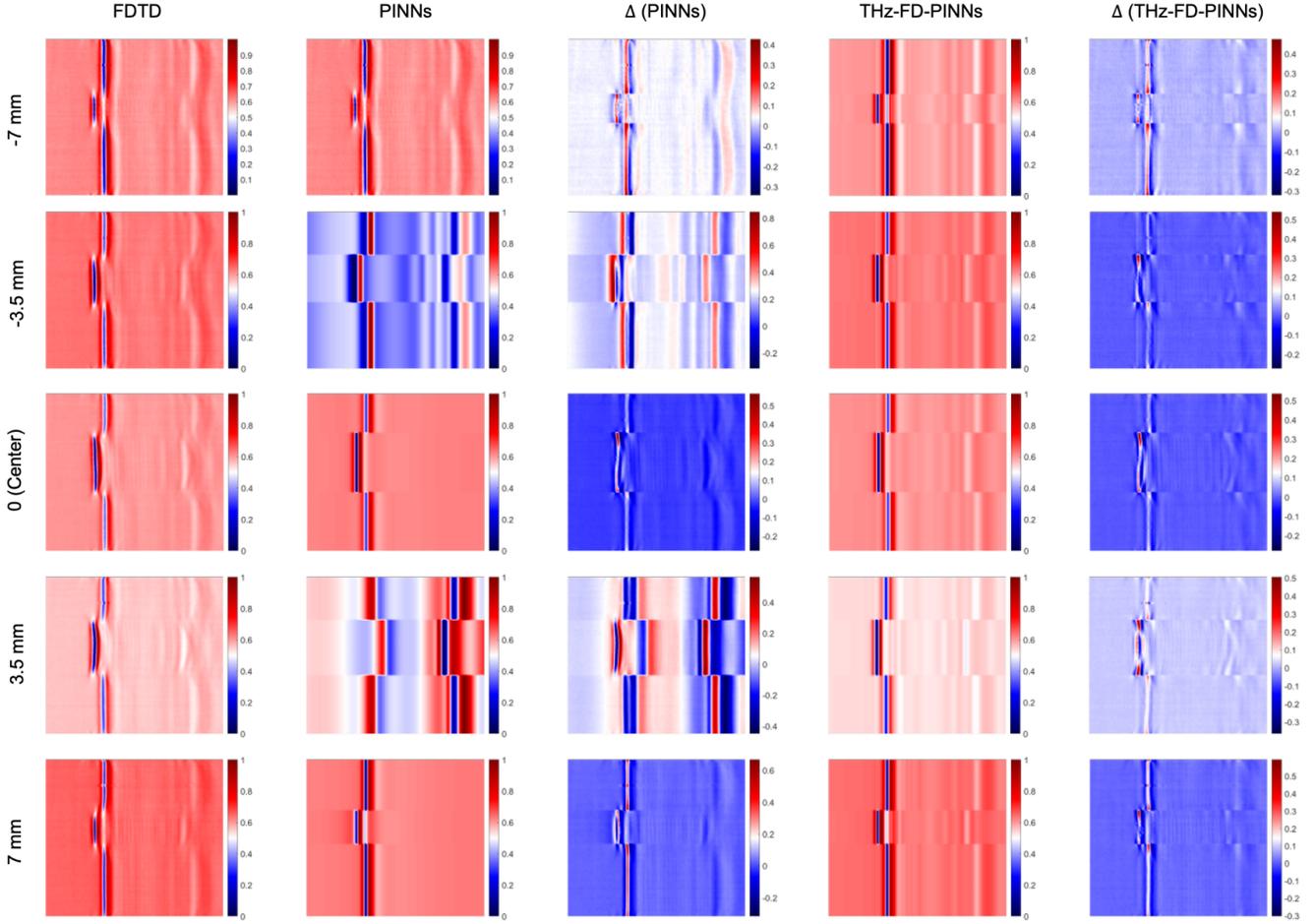

Fig. 8. Experimental results of THz-TDS, (time-domain) PINNs and THz-FD-PINNs for the spruce wood with holes. 0 denotes the center of the hole, 3.5 mm and 7 mm indicates the distance deviating to the right from the center point, -3.5 mm and -7 mm indicates the distance deviating to the left from the center point. Δ (PINNs) and Δ (THz-FD-PINNs) denote the error of PINNs and THz-FD-PINNs.

## Appendix

The time-domain physics-informed neural networks (TD-PINNs) in THz have similar network structures. However, in TD-PINNs, the loss functions are quite different. Firstly, the PDE loss term is govern by the time-domain Maxwell's equation in non-dispersive medium:

$$\mathcal{L}_{PDE} = \frac{1}{N_f}\sum_{j=1}^{N_f}(\nabla^2 \mathbf{E} - \frac{1}{c^2}\frac{\partial^2 \mathbf{E}}{\partial t^2})^2 \qquad (25)$$

For dispersive medium, the PDE loss term can be given as:

$$\mathcal{L}_{PDE} = \frac{1}{N_f}\sum_{j=1}^{N_f}(\nabla^2 \mathbf{E} - \mu\epsilon\frac{\partial^2 \mathbf{E}}{\partial t^2} - \mu\sigma\frac{\partial \mathbf{E}}{\partial t})^2 \qquad (26)$$

The boundary loss term can be given as:

$$\mathcal{L}_{BC} = \frac{1}{N_{BC}}\sum_{j=1}^{N_{BC}}(\mathbf{E} - \mathbf{f})^2 \qquad (27)$$

The initial loss term can be given as:

$$\mathcal{L}_{IC} = \frac{1}{N_{IC}}\sum_{i=1}^{N_{IC}}(|\mathbf{E}(x^i,y^i,0)|^2 + \left|\frac{\partial \mathbf{E}(x^i,y^i,0)}{\partial t}\right|^2) \qquad (28)$$

The data loss term is the same to Eq. (13). Therefore, the total loss function is:

$$\mathcal{L} = \mathcal{L}_{PDE} + \mathcal{L}_{BC} + \mathcal{L}_{IC} + \mathcal{L}_{Data} \qquad (29)$$

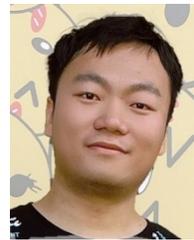

**Pengfei Zhu** received the B.Eng. degree in engineering mechanics from North University of China, Taiyuan, China, in 2019, and the M.Eng. degree in solid mechanics from Ningbo University, Ningbo, China, in 2022. He is currently working toward the Ph.D. degree in electrical engineering with Université Laval, Québec, Canada. He is a student member of IEEE, ASME, SPIE, ASNT, and CINDE.

His research interests include non-destructive testing, infrared thermography, deep learning, terahertz time-domain spectroscopy, and photothermal coherence tomography.

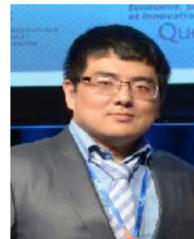

**Hai Zhang** is a full professor at Harbin Institute of Technology, Harbin, China. He received the Ph.D. degree in electrical engineering from Laval University, Quebec, QC, Canada, in 2017. He was a Postdoctoral Research Fellow with the Department of Mechanical and Industrial Engineering, University of Toronto, Toronto, ON, Canada. He was also a Visiting Researcher in Fraunhofer EZRT, Fraunhofer IZFP and Technical University of Munich, Germany. He has authored or coauthored more than 150 technical papers in peer-reviewed journals and international conferences.

He is also an Associate Editor for Infrared Physics and Technology, Measurement, and Quantitative InfraRed Thermography Journal. His research interests include nondestructive testing, industrial inspection, machine learning, medical imaging, infrared, and terahertz spectroscopy.

> REPLACE THIS LINE WITH YOUR PAPER IDENTIFICATION NUMBER (DOUBLE-CLICK HERE TO EDIT) <    10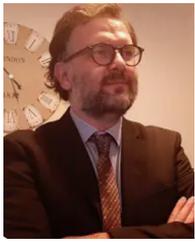

**Stefano Sfarra** received the Ph.D. degree in mechanical, management, and energy engineering from the University of L'Aquila (UNIVAQ), L'Aquila, Italy, in 2011. He worked as a Post-Doctoral Fellow at UNIVAQ until October 2017, where he became a Researcher in a fixed-term contract at the Department of Industrial and Information Engineering and Economics (DIIIE), UNIVAQ. Currently, he is an Associate Professor at DIIIE-UNIVAQ and an Adjunct Professor at Université Laval, Quebec, QC, Canada.

  He is specialized in infrared thermography, heat transfer, speckle metrology, holographic interferometry, near-infrared reflectography, energy saving, and finite element simulation techniques. Concerning these research topics, he has published more than 200 papers in journals and international conferences.

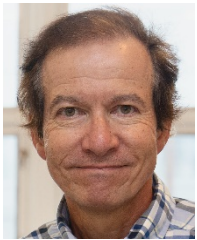

**Xavier Maldague** P.Eng., Ph.D. is full professor at the Department of Electrical and Computing Engineering, Université Laval, Québec City, Canada. He has trained over 50 graduate students (M.Sc. and Ph.D.) and contributed to over 400 publications. His research interests are in infrared thermography, NonDestructive Evaluation (NDE) techniques and vision / digital systems for industrial inspection. He is an honorary fellow of the Indian Society of Nondestructive Testing, fellow of the Canadian Engineering Institute, Canadian Institute for NonDestructive Evaluation, American Society of NonDestructive Testing. In 2019 he was bestowed a Doctor Honoris Causa in Infrared Thermography from University of Antwerp (Belguim).